\documentclass[twocolumn,
preprintnumbers,prl,nofootinbib]{revtex4}

\usepackage{amsmath,amssymb}
\usepackage{hyperref}
\usepackage{graphicx}
\usepackage{times}
\usepackage{dcolumn}

\def\vev#1{\langle#1\rangle}

\def\vol{\mathop{\mathrm{vol}}}
\def\cT{\mathcal{T}}
\def\bZ{\mathbb{Z}}

\def\Re{\mathop{\mathrm{Re}}}
\def\SU{\mathrm{SU}}
\def\UU{\mathrm{U}}
\def\SL{\mathrm{SL}}
\def\Nequals#1{$\mathcal{N}{=}\,#1$}

\begin{document}

\title{Smallest 3d hyperbolic manifolds via simple 3d theories}
\preprint{IPMU-17-0086}

\author{Dongmin Gang}
\affiliation{Center for Theoretical Physics, Seoul National University, Seoul 08826, Korea}
\author{Yuji Tachikawa}
\author{Kazuya Yonekura}
\affiliation{Kavli Institute for the Physics and Mathematics of the Universe, \\
 University of Tokyo,  Kashiwa, Chiba 277-8583, Japan}

\date{\today}

\begin{abstract}
We provide strong pieces of evidence that the mathematics of the three-dimensional hyperbolic manifolds of the first, second and third smallest volume is captured by the physics of the three-dimensional theories composed of a complex boson and a Dirac fermion, both of unit charge, coupled to a U(1) gauge field with the Chern-Simons level $-5/2$, $-7/2$ and $-3/2$, respectively.
\end{abstract}

\pacs{}
\maketitle

\section{Introduction}
Three-dimensional conformal field theories have played important roles both in condensed-matter physics and in theoretical high-energy  physics.
One class of examples is obtained by considering bosons and/or fermions coupled to Chern-Simons gauge fields.
For a long time these systems have been studied independently on the condensed-matter side and on the theoretical high-energy physics side, but recently we are seeing convergence of the two approaches, see e.g.~\cite{Ponte:2012ru,Grover:2013rc,Seiberg:2016gmd}.

In~\cite{Ponte:2012ru,Grover:2013rc}, it was argued that there is a possibility of realizing a 3d \Nequals2 supersymmetric conformal theory consisting of a single interacting chiral multiplet $\Phi$, composed of a complex scalar $\phi$ and a Dirac fermion $\psi$, on the boundary of a topological material.
In this note, we consider a slightly different system where we have a single chiral multiplet $\Phi$ of charge $+1$, coupled to an Abelian U(1) Chern-Simons gauge field with level $n{+}1/2$, where $n$ is an integer. 
These are one of the simplest non-trivial 3d conformal field theories one can write down, and belong to the class of theories studied in~\cite{Seiberg:2016gmd}.

Let us now recall that the study of supersymmetric quantum field theories has uncovered a series of unexpected relationship to various fields of mathematics. 
One such relationship exists between 3d supersymmetric theories and 3d hyperbolic manifolds, pioneered by Dimofte, Gaiotto and Gukov \cite{Dimofte:2011ju}.
To appreciate the relationship, 
let us quickly recall the bare basics of the geometry of 3d hyperbolic manifolds.

When a 3d manifold is hyperbolic, i.e.~if it admits a metric of constant negative curvature, Mostow's rigidity theorem guarantees that the metric is unique, once normalized so that $ R_{mn}= -2g_{mn} $. 
Therefore it is meaningful to talk about the volumes  as  a feature of 3d manifolds themselves.
The volumes are known to take only discrete values, and the three smallest ones are reproduced in Table~\ref{table:volume}.
These data are taken from the paper \cite{HodgsonWeeks}; they can also be obtained from the software SnapPy \cite{SnapPy}.

\begin{table*}
\centering
\begin{tabular}{|r||c|rr|c|ccc|}
\hline
& Name &  Dehn filling &&Volume  & Geodesics && \\
\hline
\hline
smallest & Weeks  & $5\mu_1-\lambda_1$, & $5\mu_2-2\lambda_2$  & 0.94271  
& 0.58460 &  0.79413 & 1.28985 \\
\hline
2nd smallest & Thurston  & $\mu_1+\lambda_1$, & $5\mu_2+\phantom{2}\lambda_2$  & 0.98136 
& 0.57808 & 0.72159   & 0.88944 \\
\hline
3rd smallest &  & $-6\mu_1+\lambda_1$, & $-3\mu_2+2\lambda_2$ & 1.01494  
&0.83144 & 0.86255 &   1.31696 \\
\hline
\end{tabular}
\caption{Data of three smallest hyperbolic manifolds.
We list for each of the manifold: common names if available,
how to obtain them by Dehn-filling the whitehead link complement, 
and lengths of the geodesics from the shortest to the third.
\label{table:volume}}
\end{table*}

\begin{figure}
\centering
\includegraphics[width=.4\textwidth]{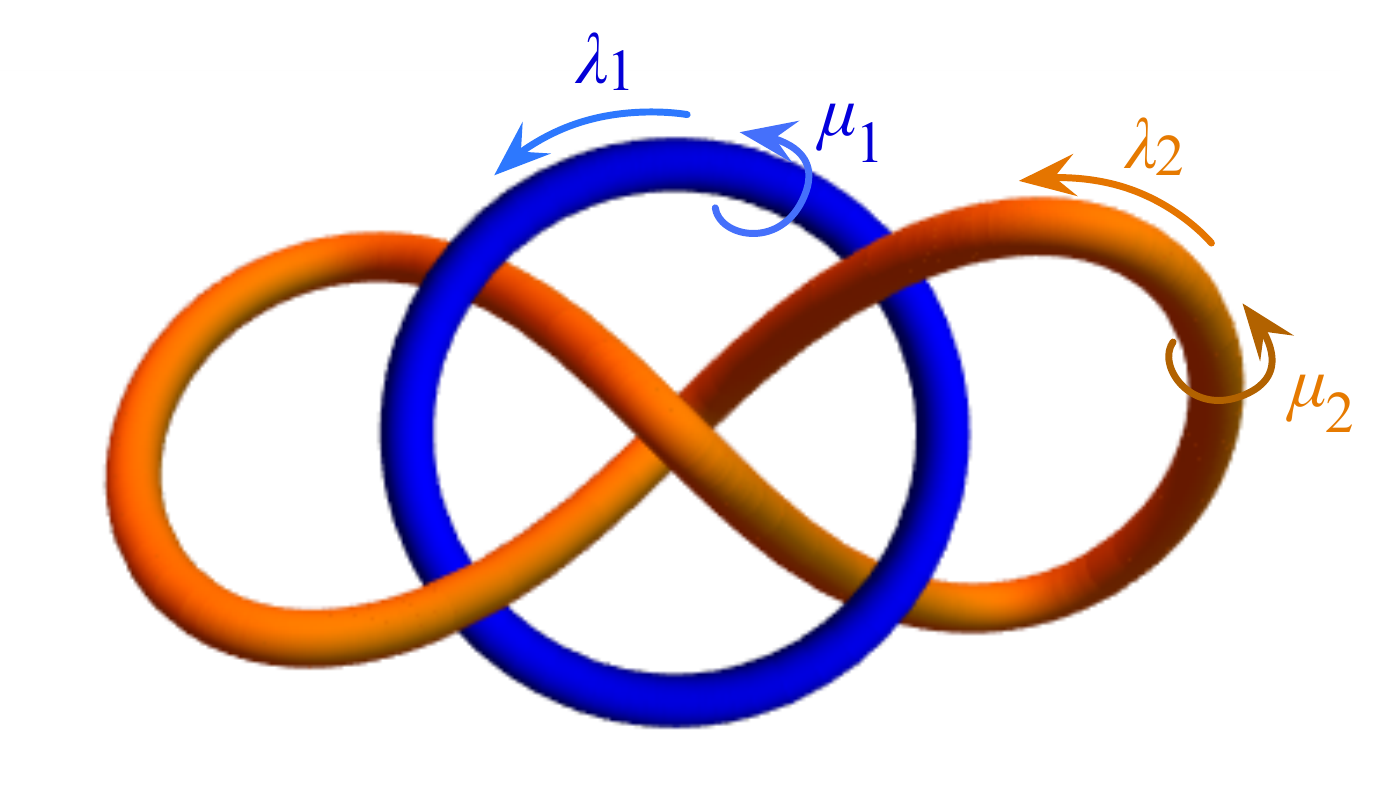}
\caption{The whitehead link, from which three smallest hyperbolic manifolds can be obtained by the Dehn filling.  \label{fig:whitehead}}
\end{figure}

Manifolds with these volumes have also explicitly been constructed. 
One method to obtain them is to start from the Whitehead link, consisting of two intertwined circles shown in Figure~\ref{fig:whitehead}.
We regard this link to be inside the 3d sphere $S^3$,
and we remove tubular neighborhoods of two circles.
The complement of the link has two boundary tori, one with the basis  $\mu_1$, $\lambda_1$ and another with $\mu_2$, $\lambda_2$, also shown in Figure~\ref{fig:whitehead}.
We now pick one linear combination of the basis for each tori, and fill them in.
This procedure is known as the Dehn filling, and results in a closed 3d manifold.
We also listed the specific linear combinations to be filled in in Table~\ref{table:volume}.

In \cite{Dimofte:2011ju}, the compactification of the 6d \Nequals{(2,0)} theory of type SU(2) on a 3d hyperbolic manifold $M$ was considered. 
From a general reasoning, the result should be  a 3d \Nequals2 superconformal theory which is denote as $T[M]$.
A basic feature of $T[M]$ is that its partition function $Z_b$ on the squashed three-sphere \begin{equation}
b^2(x^2+y^2)+\frac1{b^2} (z^2+w^2)=1
\end{equation} should behave as \begin{equation}
\Re \log Z_b \sim -\frac{1}{2\pi  b^2} \vol(M)\label{limitrelation}
\end{equation} when the squashing parameter goes to zero, $b\to 0$.

In \cite{Dimofte:2011ju}, a method to write down a 3d Lagrangian describing $T[M]$ was given for a large subclass of 3d hyperbolic manifolds, and the relation \eqref{limitrelation} was checked in many cases.
This subclass included the complement of the Whitehead link shown in Figure~\ref{fig:whitehead}, but did not include the three ones with smallest volumes tabulated in Table~\ref{table:volume}, mainly because the operation on the field theory side corresponding to the Dehn filling on the geometric side was not known.

Using an improved understanding of the implementation of the Dehn filling \cite{Bae:2016jpi,ToAppear}, we propose that the 3d superconformal theories corresponding to the 3d hyperbolic manifolds with three smallest volumes given in~Table~\ref{table:volume} are
\begin{align}
T[\text{Weeks}] &= \cT_{-5/2}^{X=0},  \label{weeks}\\
T[\text{Thurston}] &= \cT_{-7/2}^{X=1},  \\
T[\text{3rd smallest}] &= \cT_{-3/2}^{X=0},
\end{align} 
where for the ease of notation we use the symbol $\cT_{n+1/2}$ for the theory of a charged chiral multiplet coupled to a U(1) Chern-Simons gauge field at level $n+1/2$, and $X$ is an additional parameter specifying the choice of the R-symmetry, which will be explained soon below.

In the rest of this letter, we provide many pieces of evidence for this identification, including the verification of the relation \eqref{limitrelation} between the volumes and the limiting behavior of the partition functions. 
In addition, we derive the equality \eqref{weeks} for the Weeks manifold, i.e.~the smallest hyperbolic manifold, by combining the construction of \cite{Dimofte:2011ju}, the improved understanding of the Dehn filling, and known various dualities of 3d theories.

\section{Numerical evidence}
Let us first discuss the partition function  of the theory $\cT_{-n-1/2}$ on $S^3_b$. 
In general, an \Nequals2 supersymmetric theory can be placed on $S^3_b$ when a conserved R symmetry is specified. 
This R-symmetry does not have to be the R-symmetry in the superconformal algebra.
The theory $\cT_{-n-1/2}$ has a one-parameter family of conserved R-symmetries and hence the partition function depends on a free parameter 
which we denote as $X$.
The partition function is given by a by-now standard localization formula \cite{Hama:2011ea}: \begin{equation}
Z_b[X]:=\int \frac{dY}{2\pi} \Psi_b(Y) e^{-\frac{n}{4\pi i}Y^2 + \frac12(b+\frac1b) XY}
\end{equation}  where $\Psi_b$ is the quantum dilogarithm function, which behaves in the $b\to 0$ limit as
\begin{equation}
 \log \Psi_b (\frac{1}b \Upsilon )  \xrightarrow{b\to 0} \frac{1}{2\pi i b^2} \psi(e^{-\Upsilon})
\end{equation} where $\psi=\mathrm{Li}_2$ is the standard dilogarithm, and $\Upsilon=bY$ is kept fixed in the limit.

Therefore, the $b\to 0$ behavior of $Z_b[X]$ is controlled by the saddle point $\Upsilon=\Upsilon_*$ of the function \begin{equation}
W_{n,X}(\Upsilon)=\psi(e^{-\Upsilon}) - \frac{n}{2} \Upsilon^2 + \pi i X \Upsilon
\end{equation} in the following way:
\begin{equation}
\log Z_b \sim \frac{1}{2\pi i b^2} W_{n,X}(\Upsilon_*).
\end{equation}
This has the following interpretation. In the $b \to 0$ limit, the geometry becomes $S^3_b \to S^1 \times {\mathbb R}^2$ where the volume of ${\mathbb R}^2$
is regularized to be of order $b^{-2}$.
Then $W_{n,X}(\Upsilon)$ is identified as the twisted superpotential of a twisted chiral field $\Upsilon$ constructed from the $\UU(1)$ gauge multiplet in ${\mathbb R}^2$.
The terms $ \psi(e^{-Y})-  \frac{n}{2} \Upsilon^2$ come from the matter one-loop contribution and the classical Chern-Simons action,
and the term $ \pi i X \Upsilon$ can be interpreted as a coupling of the dynamical field $\Upsilon$ with a background flavor symmetry holonomy $\pi i X$ around $S^1$.
From the point of view of the 6d \Nequals(2,0) theory from which the 3d-theory/3-manifold correspondence follows, 
there is no manifest flavor symmetry and it is natural to take the flavor holonomy to be trivial.
Due to subtle reasons explained in \cite{ToAppear}, a trivial flavor holonomy corresponds not only to $e^{\pi i X} = 1$ but also to $e^{\pi i X}=-1$. 
So it is natural to take $X \in {\mathbb Z}$
to compare with the expectations from the 6d \Nequals(2,0) theory.

For the three cases above, the saddle point is respectively at \begin{equation}
\Upsilon_{*}\sim -0.141+0.704i,\ 
0.061+1.335i,\ 
+1.0472 i.
\end{equation} Then the quantity $\Re iW_{n,X}(\Upsilon_{*})$ evaluates to
\begin{equation}
\Re iW_{n,X}(\Upsilon_{*})=\begin{cases}
0.94207 & \text{for $n=2$, $X=0$};\\
 0.98137 & \text{for $n=3$, $X=1$};\\
1.01494 & \text{for $n=1$, $X=0$}.
\end{cases}
\end{equation}
which reproduces $\vol(M)$ for the Weeks, the Thurston and the third smallest manifold,
according to the relation~\eqref{limitrelation}.

We can also reproduce the lengths of the geodesics. 
In general, under the correspondence, a geodesic in the hyperbolic manifolds corresponds to a supersymmetric loop operator  $L$ in the 3d Chern-Simons-matter theories, with the relation \begin{equation}
\vev{L} = e^{\ell/2}+e^{-\ell/2}
\end{equation} in the $b\to 0$ limit, where $\ell$ is the complex length of the geodesic whose real part measures its  length.
A standard supersymmetric loop operator in the Abelian Chern-Simons-matter theory $\cT_{-n-1/2}$ is given by considering the superposition of a Wilson loop of electric charge $q$ and a vortex loop of magnetic charge $m$. 
Its vacuum expectation value can be computed using localization \cite{Kapustin:2012iw,Drukker:2012sr}, and   in the $b\to 0$ limit it is given by the formula \begin{equation}
\vev{L_{q,m}} = e^{q\Upsilon_*} (1-e^{-\Upsilon_*})^{m}.
\end{equation} 
By a straightforward computation, we see that the three shortest geodesics for the Weeks manifold are given e.g. by $(q,m)=(1,0)$, $(-1,0)$, $(-3,0)$.
An analogous analysis can also be carried out for the Thurston manifold and the third smallest manifold.

In this letter we focused only on the $b\to 0$ limit of the $S^3_b$ partition function, but we can also study its finite $b$ behavior and/or the partition function on $S^2\times S^1$, which correspond to the $\mathrm{SL}(2,\mathbb{C})$ Chern-Simons theory on the hyperbolic manifolds. 
The detail will be presented elsewhere \cite{ToAppear}.

\section{Derivation for the Weeks manifold}
Let us now give a derivation that $T[\text{Weeks}]=\cT_{-5/2}$. 
We first recall the general structure.

\paragraph{Review of the general construction:}
Dimofte, Gaiotto and Gukov \cite{Dimofte:2011ju} assigned 3d \Nequals2 theories $\cT^\text{DGG}[M]$ for a subclass of a 3d hyperbolic manifolds $M$ with a number of  $T^2$ boundaries.
One way to specify such a manifold is to start from another closed manifold $N$ with a specified link $L$. 
We then remove tubular neighborhoods of $L$ from $N$ and the result is the manifold $M=N\setminus L$: each $S^1$ component of $L$ becomes a $T^2$ boundary.
$N$ can be reconstructed from $M$ by picking a one-cycle and filling that direction for each boundary $T^2$.
Generically, $\cT^\text{DGG}[M]$ has one $\UU(1)$ flavor symmetries per each $T^2$ boundary,
and the change in the choice of the basis of $T^2$ is described by Witten's $\mathrm{SL}(2,\bZ)$ action for theories with $\UU(1)$ flavor symmetry \cite{Witten:2003ya}. 

Another way to associate 3d \Nequals2 theories to 3d hyperbolic manifolds with specified links is as follows, which will be detailed in \cite{ToAppear}.
Given a closed  3d manifold $N$ and a specified link $L$ in it, we compactify the 6d \Nequals{(2,0)} theory of type SU(2) on $N$, with the standard codimension-2 defect wrapped in $L$.
With a suitable partial topological twist, this gives rise to a 3d \Nequals2 theory we denote by $\cT^\text{6d}[N,L]$.
This theory should have one $\SU(2)$ flavor symmetry per each component of the link $L$,

When $N,L$ and $N',L'$ are obtained by a different Dehn-filling of a single 3d manifold $M$ with torus boundaries, $\cT^\text{6d}[N,L]$ and $\cT^\text{6d}[N',L']$ are related by the $\mathrm{SL}(2,\bZ)$ action for theories with $\SU(2)$ flavor symmetry by Gaiotto and Witten \cite{Gaiotto:2008ak}.
Furthermore, $\cT^\text{6d}[N]$ can be obtained from $\cT^\text{6d}[N,L]$ by the following operation.
Each $\SU(2)$ associated to a component of $L$ may have an associated chiral scalar operator $\mu$ in the adjoint, and we give a nilpotent vacuum expectation value to it.
This operation generalizes the by-now standard operation of the closure of the puncture in the 4d class S theory.

The theory $\cT^\text{DGG}[N\setminus L]$ is obtained from $\cT^\text{6d}[N,L]$ by adding to the superpotential a term $\mu^3$, the third component of the adjoint scalar, for each $\SU(2)$ symmetry.
This breaks each $\SU(2)$ to the Cartan $\UU(1)$, for generic choices of $N$ and $L$.
But it sometimes happens that $\cT^\text{DGG}[N\setminus L]=\cT^\text{6d}[N,L]$ when this superpotential deformation is not possible due to the absence of the chiral operator $\mu$.
A tell-tale sign is that the generically expected $\UU(1)$ symmetry of  $\cT^\text{DGG}[N\setminus L]$  is in fact $\SU(2)$.

\paragraph{Construction for the Weeks manifold:}
The Weeks manifold is obtained by performing the Dehn-filling of the Whitehead knot complement along $5\mu_1-\lambda_1$ for the first torus boundary and then $5\mu_2-2\lambda_2$ for the second torus boundary.
The 3d manifold after the first step is known as `the sister of the figure-eight knot complement' or as the `census manifold m003'.
This is a manifold for which the construction of Dimofte, Gaiotto, Gukov is readily applicable,
and we find \begin{equation}
\cT^\text{DGG}[\text{m003}]=\frac{\text{Two chirals $\Phi$ of charge $+1$}}{\UU(1)_0}
\end{equation} where the notation $/G_k$ means that we couple a vector multiplet of group $G$ with the Chern-Simons level $k$.
It turns out \cite{ToAppear} that this theory has an enhanced $\SU(3)$ symmetry. Up to Weyl reflections of $\SU(3)$,
the $\UU(1)$ symmetry associated to the remaining single torus boundary, generically expected for the theories constructed following Dimofte, Gaiotto, Gukov, is the Cartan of the $\SU(2) \subset \SU(3)$ acting on two chiral multiplets.
Therefore this is exactly the case where $\cT^\text{DGG}=\cT^{6d}$.

Carefully studying the conventions, we find that this $\SU(2)$ symmetry acting on two chirals is the one associated to the basis $2\mu_2-\lambda_2$.
Applying the $\SL(2,\bZ)$ action for $\SU(2)$ flavor symmetry, we find that Dehn-filling of the direction $p(2\mu_2-\lambda_2)+(\mu_2-\lambda_2)$ is given by the gauging this $\SU(2)$ symmetry of the two chirals with level $p+1/2$.
To obtain the Weeks manifold we need to choose $p=-3$. 
Then we find the following: \begin{equation}
T[\text{Weeks}]=\frac{\text{Two chirals  of charge $+1$}}{ \UU(1)_0\times \SU(2)_{-5/2}}.
\label{bar}
\end{equation}

\paragraph{Applications of known dualities:}
We now invoke one of the dualities discussed by Aharony and Fleischer \cite{Aharony:2014uya}, which states \begin{equation}
\frac{\text{Two chirals}}{\SU(2)_{-5/2}}
= \frac{\text{One chiral }}{\UU(1)_{+3/2}}.
\end{equation}
Here the $\UU(1)$ flavor symmetry acting on the two chirals on the left hand side with charge $1$
corresponds to the topological $\UU(1)^\text{top}$ symmetry on the right hand side with charge $2$.
The reason for the charge 2 is that the $\SU(2)$ gauging on the left hand side implies that all gauge invariant operators have even charges if we assign charge $1$ to the chiral fields.

The background Chern-Simons term for this $\UU(1)$ symmetry was not determined in \cite{Aharony:2014uya}, but can be determined, whose detail will be described in \cite{ToAppear}.
One finds \begin{equation}
\frac{\text{Two chirals  of charge $+1$}}{ \UU(1)_0\times \SU(2)_{-5/2}}
= \frac{\text{One chiral }}{\UU(1)_{3/2} } \Bigm/  \UU(1)^\text{top}_{1}.\label{foo}
\end{equation}

Let us denote the gauge fields of $\UU(1)_{3/2}$ and $ \UU(1)^\text{top}_{1}$ on the right hand side as $a$ and $b$, respectively.
The Chern-Simions kinetic terms on the right hand side of \eqref{foo} is described as
\begin{equation}
 \left( \frac{3}{2} \right) \frac{1}{4 \pi } a da  + 2\cdot  \frac{1}{2 \pi }b da +\frac{1}{4 \pi } b db 
\end{equation}
up to supersymmetric completion. The factor of $2$ in the second term comes from the fact that the $b$ is coupled to the topological current of $a$ by charge 2.
By shifting $b \to b - 2a$ and neglecting the almost trivial theory $\UU(1)_{1}$, we get
\begin{equation}
 \left( -\frac{5}{2} \right) \frac{1}{4 \pi } a da  .
 \end{equation}
 Therefore, the right hand side of \eqref{foo} becomes the theory of a single chiral field coupled to $\UU(1)_{-5/2}$.


\section{Acknowledgments} 
The authors would like to thank M.~Yamazaki for discussions.
YT is partially supported in part byJSPS KAKENHI Grant-in-Aid (Wakate-A), No.17H04837 
and JSPS KAKENHI Grant-in-Aid (Kiban-S), No.16H06335, and 
also supported in part by WPI Initiative, MEXT, Japan at IPMU, the University of Tokyo. The work of DG was supported by Samsung Science and
Technology Foundation under Project Number SSTBA1402-08. 
The work of KY is supported in part by the WPI Research Center Initiative (MEXT, Japan),
and also supported by JSPS KAKENHI Grant-in-Aid (17K14265). 

\bibliographystyle{ytphys}
\bibliography{ref}

\end{document}